\documentclass[twoside,twocolumn,english,prc,amsmath,a4paper,myheadings]{revtex4}
\usepackage[T1]{fontenc}
\usepackage{geometry}
\usepackage{amsmath}
\usepackage{graphicx}
\usepackage{amssymb}

\makeatletter
\def\@oddhead{\rightmark \hfill  V2 Scaling in PbPb at 2.76 TeV  \hfill \thepage}
\def\@evenhead{\thepage \hfill K. Werner\hfill}
\def\fnum@table{\tablename~{\bf\thetable}}
\def\fnum@figure{\figurename~{\bf\thefigure}}
\def\tablename{\footnotesize{\bf Table}}
\def\figurename{\footnotesize{\bf Figure}}

\usepackage{dcolumn}

\def\citet{\cite}

\makeatother

\usepackage{babel}

\begin{document}

\title{V2 Scaling in PbPb Collisions at 2.76 TeV}

\author{{\normalsize K.$\,$Werner}}

\address{SUBATECH, University of Nantes -- IN2P3/CNRS-- EMN, Nantes, France}

\begin{abstract}
We investigate scaling properties of the elliptical flow parameter
$v_{2}$ in PbPb Collisions at 2.76 TeV within a recently introduced
new theoretical scheme which accounts for hydrodynamically expanding
bulk matter, jets, and the interaction between the two. 
\end{abstract}
\maketitle
The transverse momentum ($p_{t}$) dependence of the elliptical flow
parameter $v_{2}$ for identified hadrons has been investigated in
great detail in heavy ion collisions at 200 GeV. A mass splitting
has been observed at low $p_{t}$: the proton curve for example is
shifted to the right compared to the pions, whereas at higher $p_{t}$
the curves cross, the proton result finally being 50 \% above the
pion one. If one plots $v_{2}/n_{q}$ versus $K\! E_{t}/n_{q}$ (with
$n_{q}$ being the number of quarks of the hadron and $K\! E_{t}$
the transverse kinetic energy) one observes a unique curve for many
hadron species, referred to as quark number scaling. This is usually
considered to support the idea of quark coalescence \citet{coa2,coa3,coa4,coa5}.
However, as shown in \citet{phenix}, the situation is more complex:
scaling is observed in central AuAu collisions at 200 GeV, whereas
in more peripheral collisions the scaling is broken for $K\! E_{t}>0.7$GeV.

In ref. \citet{jetbulk}, we introduced a new theoretical scheme which
accounts for hydrodynamically expanding bulk matter, jets, and the
interaction between the two. The whole transverse momentum range is
covered, from very low to very high $p_{t}$. In \citet{jetbulk},
we show that the new approach can accommodate spectra of jets with
$p_{t}$ up to 200 GeV/c in $pp$ scattering at 7 TeV, as well as
particle yields and harmonic flows with $p_{t}$ between 0 and 20
GeV/c in PbPb collisions at 2.76 TeV. Since our aim is a single model
which is able to cover all phenomena, we will apply the approach of
ref. \citet{jetbulk}, with exactly the same parameters (EPOS2.17v3),
to study the question of scaling (or not) in PbPb collisions at 2.76
TeV.

The starting point of the new approach are color flux tubes which
appear as a consequence of hard scatterings. In heavy ion collisions,
we have many of these flux tubes, which constitute eventually both
bulk matter (which thermalizes, flows, hadronizes, and finally performs
hadronic scatterings) and jets, according to some criteria based on
partonic energy loss. 

The flux tubes are treated as kinky strings, where the kinks amount
to transversely moving string pieces carrying the transverse momenta
of the hard partons. Three possibilities occur: (A) String segments
which have not sufficient energy to escape will constitute matter,
they loose their character as individual strings. This matter will
evolve hydrodynamically and finally hadronizes ({}``soft hadrons'').
(B) String segments having sufficient energy to escape and being formed
outside the matter, constitute jets ({}``jet-hadrons''). (C) There
are finally also string segments produced inside matter or at the
surface, but having enough energy to escape and show up as jets ({}``jet-hadrons'').
They are affected by the flowing matter ({}``fluid-jet interaction''). 

\begin{figure}[tb]
\begin{raggedright}
\vspace*{0.5cm}
\par\end{raggedright}

\begin{raggedright}
{\large \hspace*{1.2cm}(a)}
\par\end{raggedright}{\large \par}

\vspace*{-0.5cm}

\begin{centering}
\includegraphics[scale=0.15]{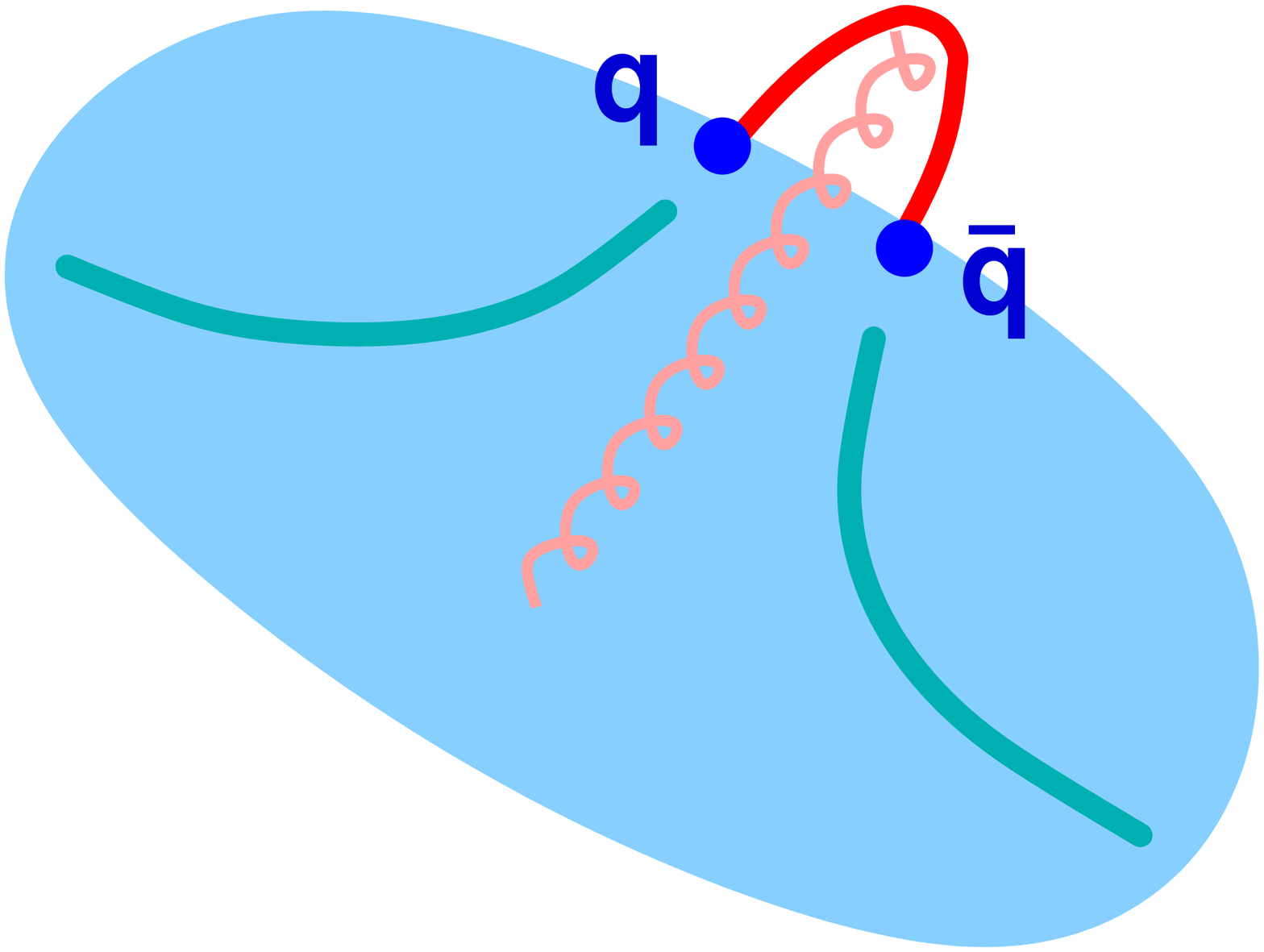}
\par\end{centering}

\begin{raggedright}
{\large \hspace*{1.2cm}(b)}
\par\end{raggedright}{\large \par}

\vspace*{-0.5cm}

\begin{centering}
\includegraphics[scale=0.15]{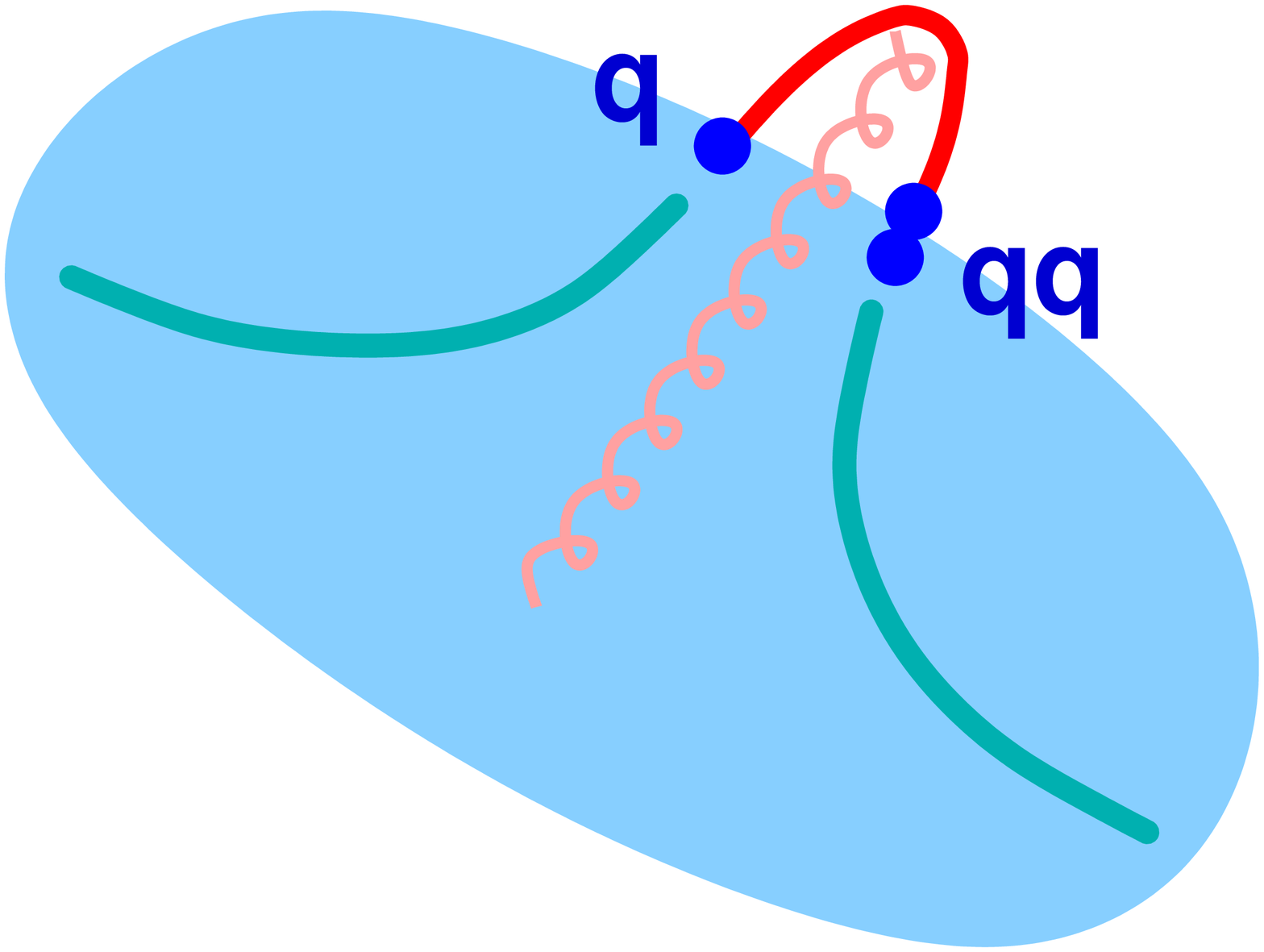}
\par\end{centering}

\caption{(Color online) Escaping string segment, getting it's endpoint partons
from the fluid. We show the case of a quark and an antiquark (a) and
of a quark and a diquark (b). The rest of the string dissolves in
matter.\label{stgfr3}}

\end{figure}

Interesting is case (C). The jet-hadrons are produced still inside
matter or at the surface, but they escape. Here we assume that the
quark, antiquark, diquark, or antidiquark needed for the flux tube
breaking is provided by the fluid with properties (momentum, flavor)
determined by the fluid rather than the Schwinger mechanism, whereas
the rest of the string dissolves in matter, see fig. \ref{stgfr3}.
Considering transverse fluid velocities up to 0.7c, and thermal parton
momentum distributions, one may get a {}``push'' of a couple of
GeV to be added to the transverse momentum of the string segment.
Important for the discussion in this paper: baryons ($n_{q}=3$) are
more pushed than mesons ($n_{q}=2$). This property is similar to
the coalescence mechanism, but there is also a substantial difference:
the sum of the transverse momenta of the $n_{q}$ quarks is only a
fraction of the total $p_{t}$, since an important contribution comes
from the flux tube segment (carrying the parton $p_{t}$). At large
$p_{t}$ the latter contribution dominates, quark number scaling will
be violated.

\begin{figure}[tb]
\begin{raggedright}
{\large \hspace*{0.8cm}(a)}
\par\end{raggedright}{\large \par}

\vspace*{-0.2cm}

\begin{centering}
\includegraphics[angle=270,scale=0.32]{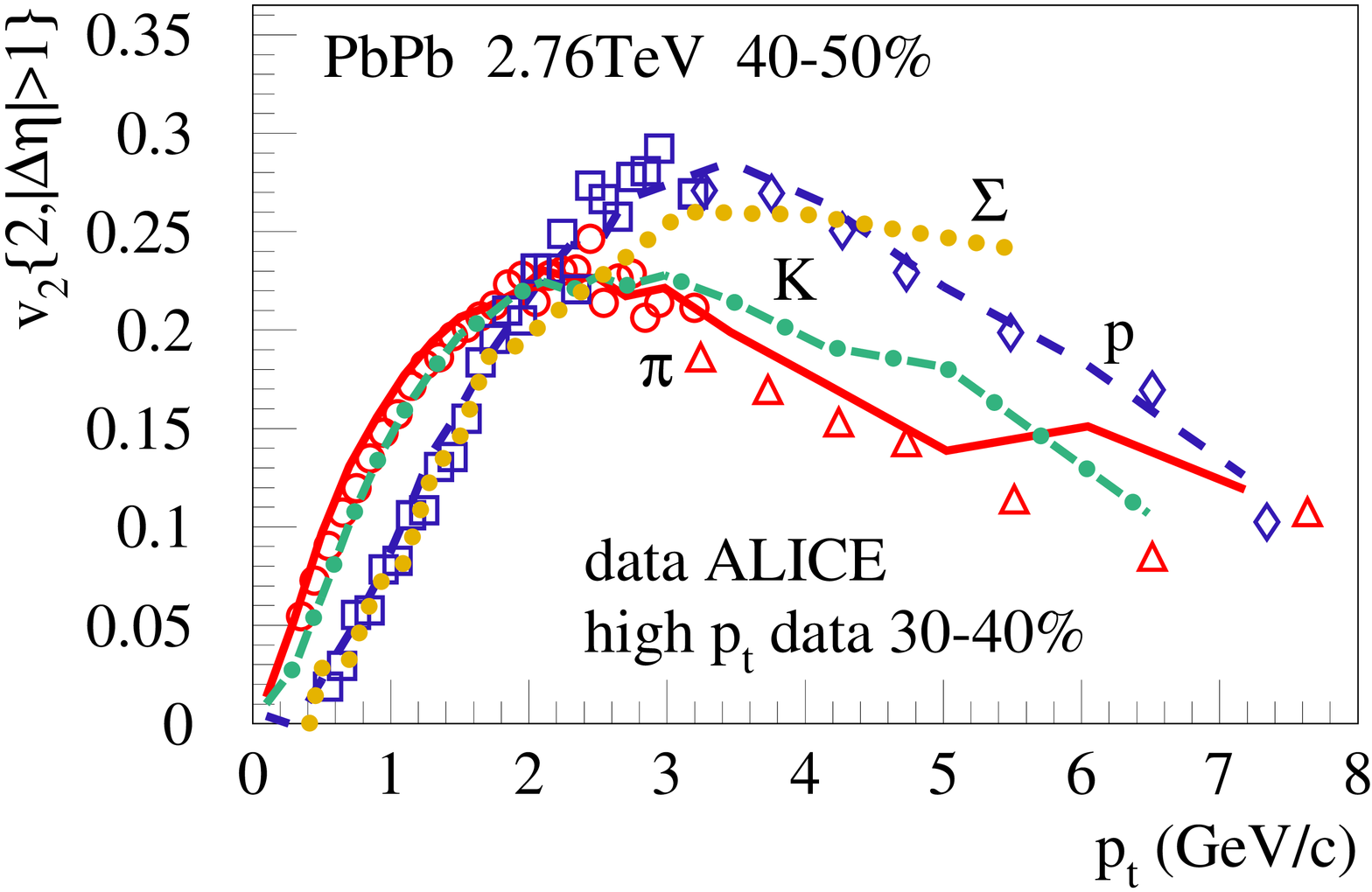}
\par\end{centering}

\begin{raggedright}
\vspace*{-0.2cm}
\par\end{raggedright}

\begin{raggedright}
{\large \hspace*{0.8cm}(b)}
\par\end{raggedright}{\large \par}

\vspace*{-0.2cm}

\begin{centering}
\includegraphics[angle=270,scale=0.32]{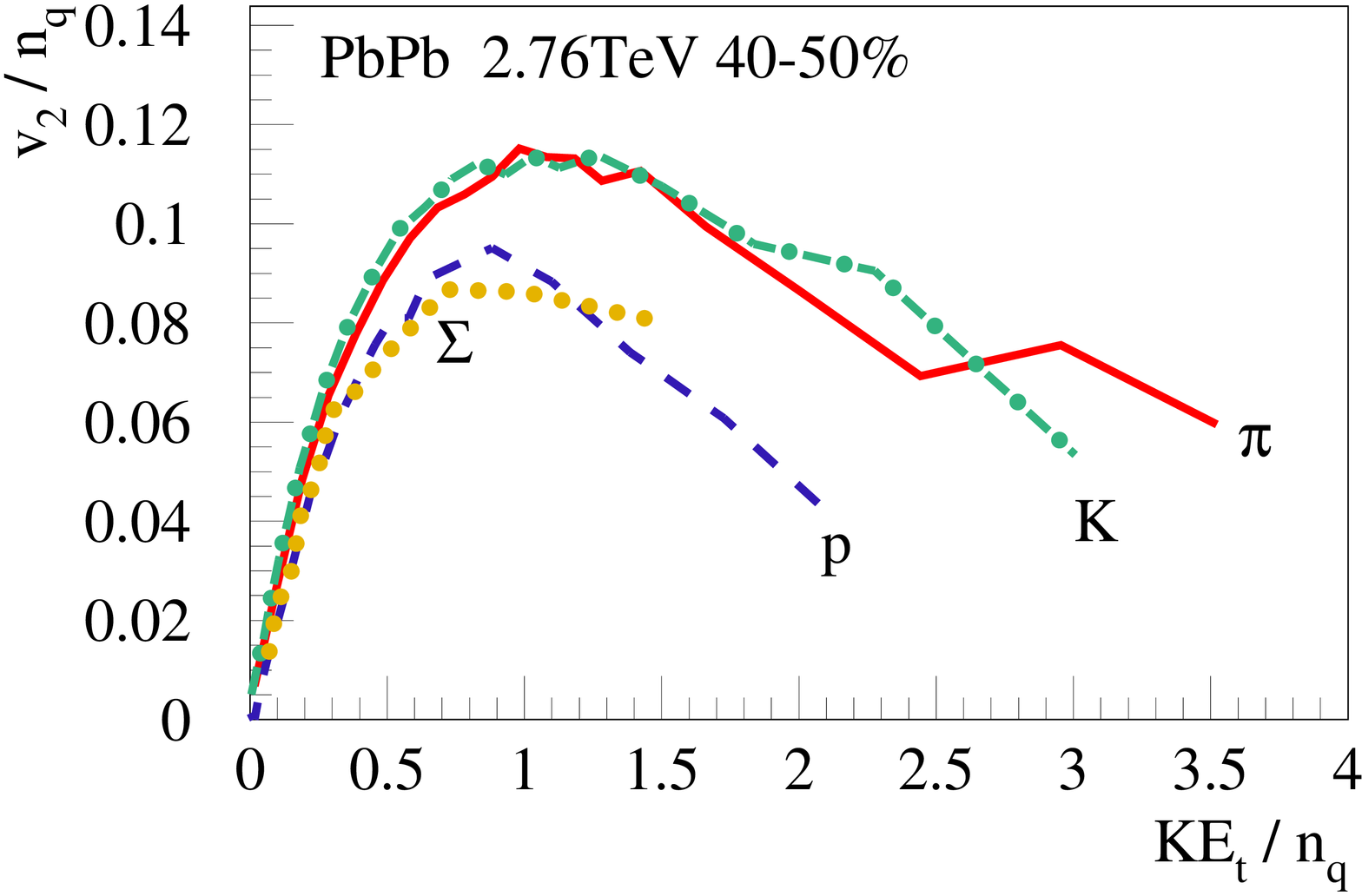}
\par\end{centering}

\caption{(Color online) (a) Transverse momentum dependence of $v_{2}$ for
pions (full line), kaons (dashed-dotted line), protons (dashed line),
and sigma baryons (dotted line) in semi-peripheral PbPb collisions
(40-50\%) at 2.76 TeV. We also show pions (circles and triangles)
and protons (squares and rhombi) from ALICE \citet{alice} (the high
$p_{t}$ data refer to 30-40\%).\label{fig:f30} (b) Scaling representation
of the curves from (a).}

\end{figure}

In fig. \ref{fig:f30}(a), %
we show the transverse momentum dependence of $v_{2}$ for pions (full
line), kaons (dashed-dotted line), protons (dashed line), and sigma
baryons (dotted line) in semi-peripheral PbPb collisions (40-50\%)
at 2.76 TeV. We also show pions (circles and triangles) and protons
(squares and rhombi) from ALICE \citet{alice} (the high $p_{t}$
data refer to 30-40\%). We clearly see the mass splitting at low $p_{t}$
and crossings at higher $p_{t}$. %
\begin{figure}[b]
\begin{raggedright}
{\large \hspace*{0.8cm}(a)}
\par\end{raggedright}{\large \par}

\vspace*{-0.1cm}

\begin{centering}
\includegraphics[angle=270,scale=0.32]{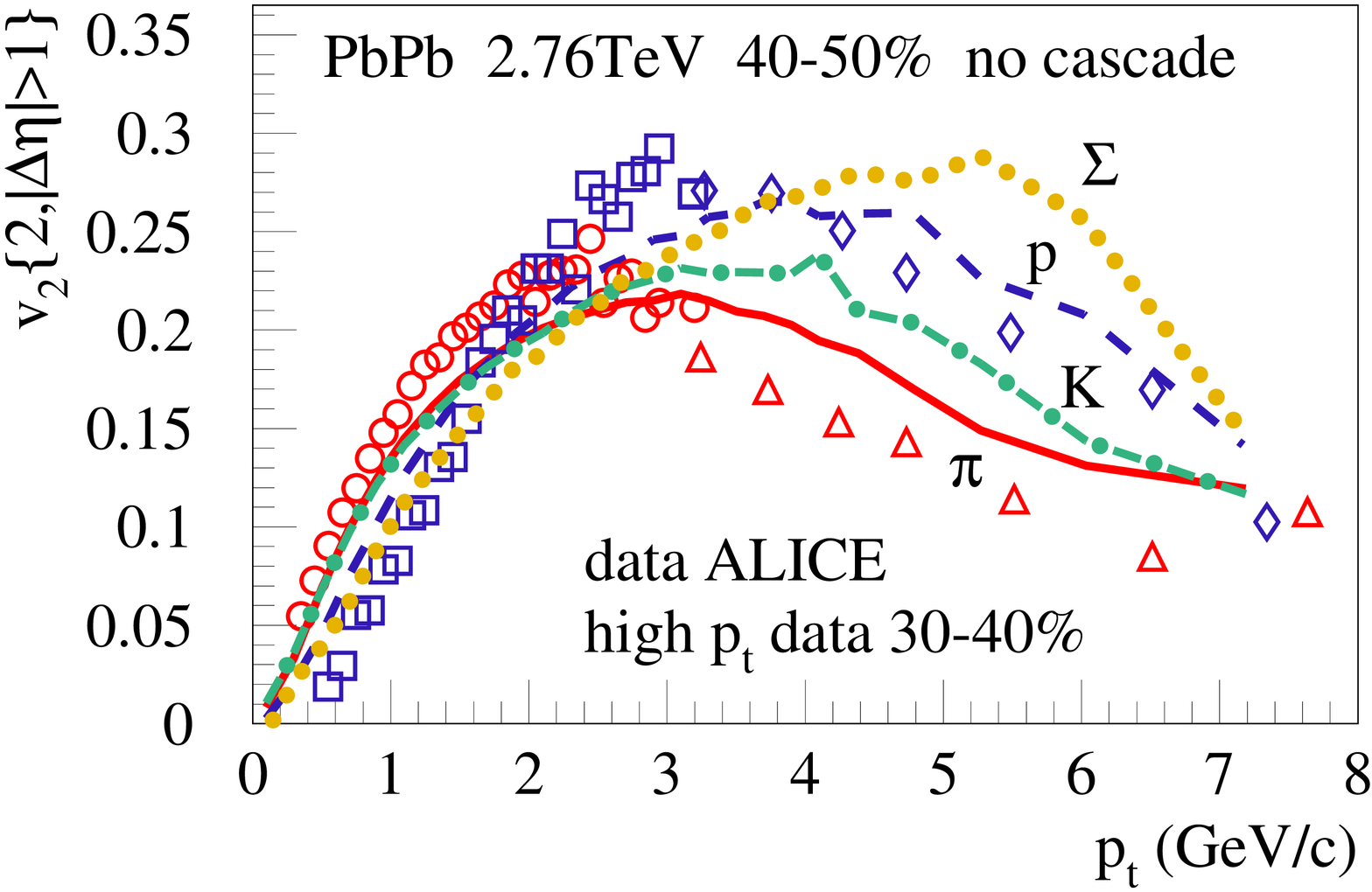}
\par\end{centering}

\begin{raggedright}
\vspace*{-0.2cm}
\par\end{raggedright}

\begin{raggedright}
{\large \hspace*{0.8cm}(b)}
\par\end{raggedright}{\large \par}

\vspace*{-0.1cm}

\begin{centering}
\includegraphics[angle=270,scale=0.32]{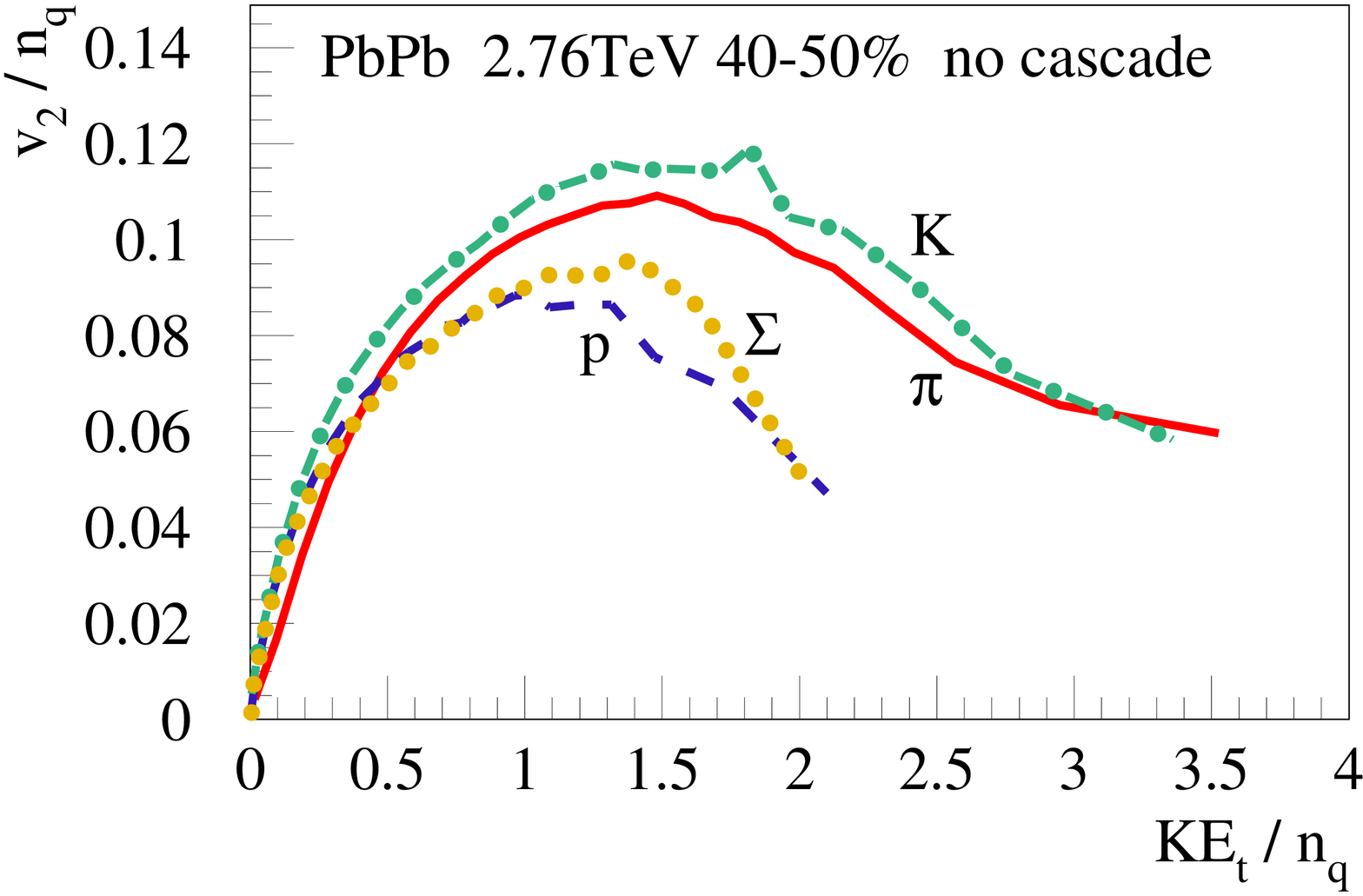}
\par\end{centering}

\caption{(Color online) Same as fig. \ref{fig:f30}, but calculation without
hadronic cascade. \label{fig:f33}}

\end{figure}
In fig. \ref{fig:f30}(b), we plot the four curves as $v_{2}/n_{q}$
versus $K\! E_{t}/n_{q}$ (scaling representation). Whereas we see
scaling at small $K\! E_{t}$, at high $K\! E_{t}$ the mesons and
baryons separate (scaling violation), which is at least very clear
for the protons, for the sigma baryons we are limited to relatively
small $p_{t}$, due to statistics. 

To understand the effect of hadronic final state rescattering, we
plot in fig. \ref{fig:f33} the corresponding curves for the calculations
without cascade. As expected (from RHIC), at low $p_{t}$ there is
much less mass splitting, and consequently there is no scaling. So
the low $K\! E_{t}$ scaling is a hadronic effect (maybe due to the
fact that $MN$ cross sections are roughly 2/3 of $BN$ cross sections
($B$ being a baryon, $M$ a meson, $N$ a nucleon). At high $K\! E_{t}$,
the results are similar to the ones from the full calculation. 

To summarize: We predict scaling only for small values of $K\! E_{t}$,
at larger values the scaling is violated, clearly seen for the protons.
We attribute this to the fact that with increasing $p_{t}$ the deviation
of our approach from simple coalescence increases.

\vfill{}


\begin{thebibliography}{1}
\bibitem{coa2}V. Greco, C. M. Ko, and P. Levai, Phys. Rev. Lett.
90, 202302 (2003); Phys. Rev. C 68, 034904 (2003).

\bibitem{coa3}R. J. Fries, B. Mueller, C. Nonaka, and S. A. Bass,
Phys. Rev. Lett. 90, 202303 (2003); Phys. Rev. C 68, 044902 (2003). 

\bibitem{coa4}R. C. Hwa and C. B. Yang, Phys. Rev. C 67, 034902 (2003);
Phys. Rev. C 70, 024905 (2004). 

\bibitem{coa5}S. Pratt, S. Pal, Phys.Rev. C71 (2005) 014905.

\bibitem{phenix}PHENIX Collaboration, K. Adare et al, arXiv:1203.2644

\bibitem{jetbulk}K.Werner, Iu.Karpenko, M. Bleicher, T.Pierog, S.
Porteboeuf-Houssais, arXiv:1203.5704.

\bibitem{alice}A. Dobrin for the ALICE collaboration, arXiv:1107.0454,
Mikolaj Krzewicki for the ALICE Collaboration, arXiv:1107.0080
\end{thebibliography}
\end{document}